# Reward prediction errors arising from switches between major and minor modes in music: An fMRI study

**Chen-Gia Tsai [1, 2], Yi-Fan Fu[3], and Chia-Wei Li [4*]**

[1] Graduate Institute of Musicology, National Taiwan University, Taipei, Taiwan
[2] Neurobiology and Cognitive Science Center, National Taiwan University, Taipei, Taiwan
[3] Department of Bio-Industry Communication and Development, National Taiwan University, Taipei, Taiwan
[4] Department of Radiology, Wan Fang Hospital, Taipei Medical University, Taipei, Taiwan

**ABSTRACT**

Evidence has accumulated that prediction error processing plays a role in the enjoyment of music listening. The present study examined listeners' neural responses to the signed reward prediction errors (RPEs) arising from switches between major and minor modes in music. We manipulated the final chord of J. S. Bach's keyboard pieces so that each major-mode passage ended with either the major (Major-Major) or minor (Major-Minor) tonic chord, and each minor-mode passage ended with either the minor (Minor-Minor) or major (Minor-Major) tonic chord. In Western music, the major and minor modes have positive and negative connotations, respectively. Therefore, the outcome of the final chord in Major-Minor stimuli was associated with negative RPE, whereas that in Minor-Major was associated with positive RPE. Twenty-three musically experienced adults underwent functional magnetic resonance imaging while listening to Major-Major, Major-Minor, Minor-Minor, and Minor-Major stimuli. We found that activity in the subgenual anterior cingulate cortex (extending into the ventromedial prefrontal cortex) during the final chord for Major-Major was significantly higher than that for Major-Minor. Conversely, a frontoparietal network for Major-Minor exhibited significantly increased activity compared to Major-Major. The contrasts between Minor-Minor and Minor-Major yielded regions implicated in interoception. We discuss our results in relation to executive functions and the emotional connotations of major versus minor mode.

## 1. INTRODUCTION

The processing of language and gestures involves dynamic interactions between cognition and emotion. Our brain generates prediction about future units by combining sensory input, contextual information, and prior knowledge (de Lange *et al.*, 2018; Siman-Tov *et al.*, 2019; Zhang *et al.*, 2021). The induced emotions depend on both the valence of outcome and the difference between prediction and outcome, namely prediction error (Villano *et al.*, 2020). Emotional responses are, in turn, essential for acquiring knowledge of temporal patterns and learning adaptive behaviors (Berridge, 2001; Haruno & Kawato, 2006; D'Astolfo & Rief, 2017).

Temporal patterns occurring in music can modulate listeners' emotions, and there is recent interest in the relationship between music-induced emotions and predictive processing. Witek *et al.* (2014) showed that rhythms of intermediate complexity evoked listeners' maximal pleasure and desire to move. This finding was explained in terms of the interaction between prediction error and the certainty (precision) of predictions (Vuust *et al.*, 2018). Gold *et al.* (2019) reported that listeners experienced most pleasure in response to music of intermediate complexity, which maximizes both reducible uncertainty and learnable information. In a study of chord progressions in pop songs, chords with low uncertainty and high surprise as well as those with high uncertainty and low surprise were found to receive high pleasure ratings (Cheung *et al.*, 2019). Koelsch *et al.* (2019) posited that musical prediction errors evoke pleasure and attract attention because they resolve the ensuing uncertainty.

Although these prior studies reached a consensus on the point that the cognitive processing of prediction error is an integral part of the enjoyment of music listening, none of them took into account the value of predicted or actual outcome. Although the value of a musical event is sometimes elusive and difficult to measure, its importance is twofold: first, listeners' emotional responses to a musical event depend on its value; second, this value allows for considering the effect of signed reward prediction error (RPE) on emotional responses. Positive RPE or pleasant surprise occurs when a reward is greater than predicted, whereas negative RPE or unpleasant surprise occurs when a reward is smaller than predicted (Schultz, 2016; 2017). To date, only limited studies have examined the neural correlates of music-induced signed RPE. In a previous study from our team, participants listened to excerpts of pop songs in which the verse was followed by either the chorus or noise. Given that a rewarding element of pop songs is the chorus preceded by the verse, the noise condition was associated with a negative RPE, which resulted in decreased activity in the ventromedial prefrontal cortex (vmPFC) and increased activity in the ventrolateral prefrontal cortex (vlPFC) and ventral anterior insula (Li *et al.*, 2015). However, the sequence of a song verse followed by noise hardly occurs in real life.

An important kind of musical idiom that may manifest the effect of signed RPE is switches between major and minor modes. For listeners familiar with Western music, the major mode has positive connotations, whereas minor-mode music often expresses sadness and is associated with affectively negative words (Hevner, 1935; Kastner & Crowder, 1990; Nieminen *et al.*, 2012; Justus *et al.*, 2018). Therefore, a switch from major to minor mode in music a negative RPE in a listener's brain. Conversely, a switch from minor to major mode presumably generates a positive RPE. The latter is best exemplified by a *Picardy cadence*, which refers to the use of a major chord of the tonic at the end of a minor-mode piece or section.

Here, we manipulated the final chord of music pieces to examine how listeners' neural activation was modulated by mode switch. There were four types of musical stimuli: a major-mode passage ending with the major tonic chord (Major-Major), a major-mode passage ending with the minor tonic chord (Major-Minor), a minor-mode passage ending with the minor tonic chord (Minor-Minor), and a minor-mode passage ending with the major tonic chord (Minor-Major). Throughout the brain-imaging experiment, stimuli of these four types were presented in a pseudorandom order and with equal probability. The outcome of the final chord in the Major-Minor stimuli was associated with a negative RPE, whereas that in the Minor-Major stimuli was associated with a positive RPE.

The aim of the present study was to probe the dynamics of neural responses to music-induced RPE using functional magnetic resonance imaging (fMRI). Prior research has demonstrated that the striatum and ventral anterior insula are sensitive to unsigned RPE, the lateral prefrontal cortices preferentially respond to negative RPE, and the vmPFC preferentially responds to positive RPE (Yacubian *et al.*, 2006; Fouragnan *et al.*, 2018). We hypothesized that the final chord in all four stimulus types would commonly activate the striatum and ventral anterior insula, the contrast of Major-Minor minus Major-Major would yield activation in the lateral prefrontal cortices, and the contrast of Minor-Major minus Minor-Minor would yield activation in the vmPFC.

## 2. MATERIAL AND METHODS

*2.1 Participants*

Twenty-three adults (age range, 20–30 years; 13 men) completed the fMRI experiment, in which three fMRI scanning runs for the present study (focusing on mode switch) alternated with two fMRI scanning runs for another study focusing on the diminished seventh chord. This design was used to minimize affective habituation that could occur with repeated exposure to the same music. The selection and recruitment of participants are described in the next paragraph. The other methods and results of the study on the diminished seventh chord are not mentioned further in this paper.

Participants were recruited via a public announcement on the internet, which stated the requirement of a high familiarity with Western classical music. In a pre-scan online test, volunteers were presented with four chord progressions and asked to judge whether a tonality change occurred in each stimulus. Regarding switches between major and minor modes, they were further presented with four music passages and asked to identify the type of each passage (i.e., Major-Major, Major-Minor, Minor-Major, or Minor-Minor). Fifty-one volunteers had a 100% correct rate in this pre-scan test and were therefore invited to participate in the fMRI experiment. Twenty-five healthy adults were enrolled in this experiment. One was found to wear a retainer on more than one tooth during the scanning session and was therefore excluded from the experiment. Another reported discomfort during the scanning session and withdrew from the experiment. Therefore, the final analyzed sample comprised 23 participants. According to self-reports, no participants had a history of neurological or psychiatric disorders. Eleven participants had studied musical instruments for 6 years or more. The participants were compensated with approximately 16 USD after the completion of this experiment. Written informed consent was obtained from each participant prior to their participation in the study. All research procedures were performed in accordance with a protocol approved by the Institutional Review Board of National Taiwan University (202007HM031). This study was conducted in accordance with the Declaration of Helsinki.

*2.2 Stimuli*

J. S. Bach's keyboard pieces were used as stimuli to enhance ecological validity. J. S. Bach and his contemporaries commonly used Picardy cadences at the ends of minor-mode works. All musical stimuli were the final passages of keyboard pieces selected from Bach's *Well-Tempered Clavier*. We downloaded the musical instrument digital interface (MIDI) files at http://www.kunstderfuge.com/bach/harpsi.htm to generate the four types of stimulus: Major-Major, Major-Minor, Minor-Major, and Minor-Minor. The Major-Major stimuli were extracted from the following works: prelude from BWV846, fugue from BWV846, fugue from BWV848, fugue from BWV850, prelude from BWV860, fugue from BWV860, prelude from BWV862, fugue from BWV862, and prelude from BWV868. The Minor-Major stimuli were extracted from the following works: fugue from BWV847, fugue from BWV851, prelude from BWV855, fugue from BWV857, fugue from BWV859, fugue from BWV861, fugue from BWV867, prelude from BWV 869, and fugue from BWV873. The Major-Minor and Minor-Minor stimuli were generated by shifting the thirds of the final chords in the Major-Major and Minor-Major stimuli a semitone down (Figure 1). There were nine musical stimuli in each stimulus type.

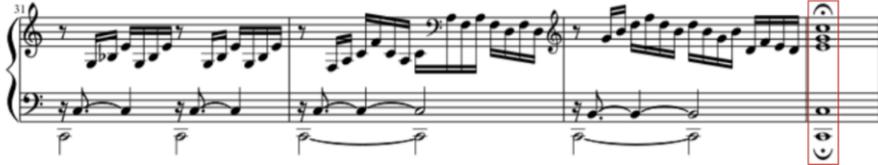
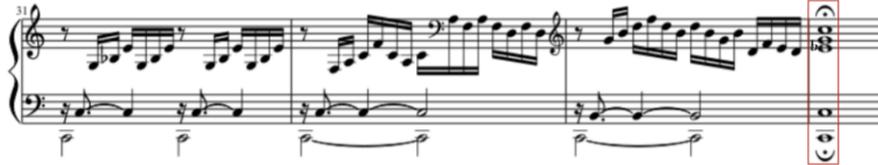
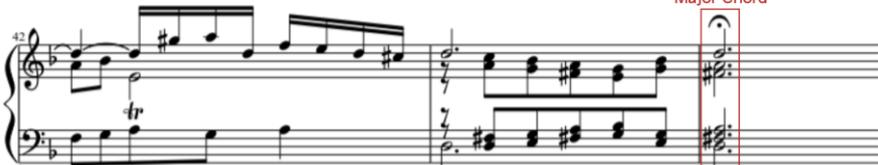
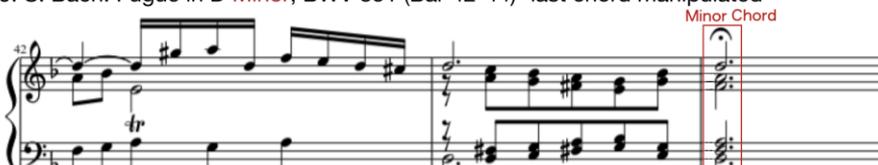

**Figure 1.** Examples of the four stimulus types.

These MIDI files were processed with MuseScore composition and notation software (Musescore BVBA, Belgium) and converted to MP3 files (44100 Hz sampling rate, 16-bit stereo) with a virtual musical instrument named "Piano". A final ritardando (gradual decrease in the tempo) was applied to each stimulus to amplify the emotional impact of the music (Friedman, 2018). All stimuli had a fade-in of 0.5 s. The duration of each stimulus was 14 s. The tempo ranged from 104 to 200 bpm.

*2.3 Procedure*

The experiment had three parts: a pre-scan practice session, a scanning session, and a post-scan questionnaire. On the day of neuroimaging, participants underwent a pre-scan practice session outside the scanner to familiarize them with the procedure. In this 3-min practice session, they were presented with four sample stimuli in a pseudorandom order, with one stimulus for one type. After the presentation of each stimulus, the participants were asked to rate subjective emotional valence on a four-point Likert scale (1 = mostly negative; 2 = somewhat negative; 3 = somewhat positive; 4 = mostly positive).

The scanning session began with an anatomical scan of the whole brain. There were three fMRI scanning runs for the present mode-switch study. Each run consisted of 24 trials and lasted 7.3 minutes. Each trial began with a silence (1.7 s) and a warning tone (2.5 kHz, 0.3 s). An auditory stimulus was presented from the beginning of the fourth second onwards. The participants were instructed to rate subjective emotional valence on a four-point Likert scale by pressing a button after hearing the final chord, the onset of which occurred between the 13.5 s and 14 s (Figure 2). The assignment of labels to the response buttons was counterbalanced across participants. Each of the 36 musical stimuli (9 stimuli × 4 types) was presented twice throughout the scanning session. The 72 trials were presented in a pseudorandom order during the scanning session. No stimuli in the pre-scan test and practice session were presented in the scanning session of this study. Throughout the scanning session, the participants were instructed to close their eyes.

In the post-scan questionnaire, the participants were presented with seven mode-switch stimuli in a pseudorandom order. They were asked to identify the type of each stimulus (i.e., Major-Major, Major-Minor, Minor-Major, or Minor-Minor). Finally, they reported their familiarity with the musical stimuli in the present study and their learning experience of any instrument.

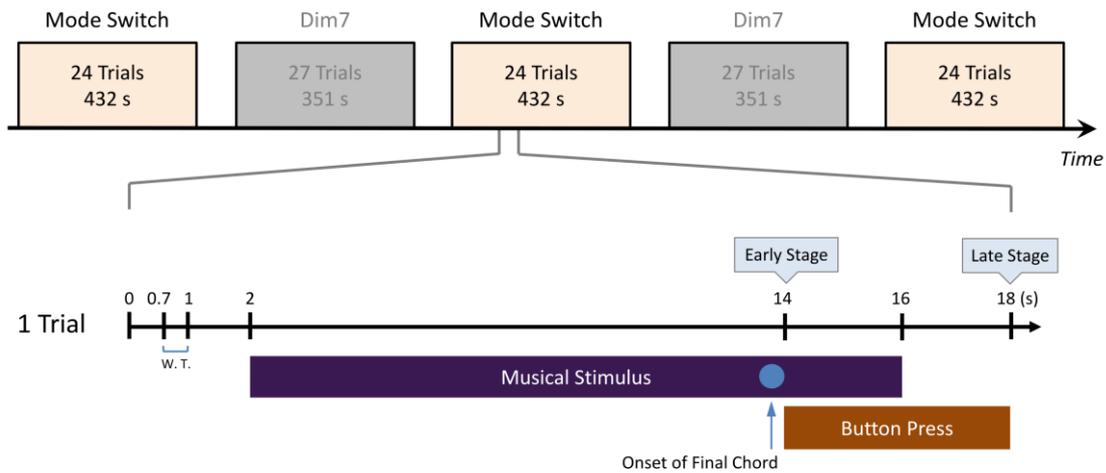

**Figure 2.** Schematic description of experimental procedure.
Abbreviation: Dim7, diminished seventh chord.

*2.4 MRI data acquisition*

Functional and structural images were acquired on a 3 Tesla MR system (MAGNETOM Prisma, Siemens, Erlangen, Germany) and a 20-channel array head coil at the Imaging Center for Integrated Body, Mind, and Culture Research, National Taiwan University. In the functional scans, axial images about 2.5 mm thick were acquired using gradient echo planar imaging (EPI) with the following parameters: time to repetition = 2500 ms, echo time = 30 ms, flip angle = 87°, in-plane field of view = 192 × 192 mm, and acquisition matrix = 78 × 78 × 45 to cover whole cerebral areas. For spatial individual-to-template normalization in preprocessing, magnetization-prepared rapid gradient echo T1-weighted images with an isotropic spatial resolution of 0.9 mm were acquired for each participant.

*2.5 Data analysis*

The data from the rating of emotional valence for each stimulus were averaged across participants. Shapiro tests were used to assess the normality of the distribution of the emotional valence rating data for each stimulus type. Subsequently, either paired *t*-tests or Wilcoxon signed-rank tests were used to examine whether there were significant differences in the emotional valence ratings between Major-Major and Major-Minor as well as between Minor-Major and Minor-Minor.

Preprocessing and statistical analysis of the fMRI data were performed using the SPM12 toolbox (Wellcome Trust Centre for Neuroimaging, London, UK). The first four volumes of each run were discarded to exclude T1 saturation effects. The functional images were corrected for differences in slice-acquisition time to the first volume and were realigned to the first volume in each scanning run using an affine transformation. Coregistered images were normalized to the standard Montreal Neurological Institute (MNI) average template and resampled to an isotropic voxel size of 2 mm. Normalized images were spatially smoothed with a Gaussian kernel of 5-mm at full width at half maximum.

The data of each participant were analyzed using the general linear model by fitting the time series data with the canonical hemodynamic response function (HRF) modeled at the events of early ($14^{th}$ second) and late ($18^{th}$ second) stages of music outcome processing. Linear contrasts were computed to characterize responses of interest, averaging across fMRI runs. The group-level analysis consisted of four paired *t*-tests for the contrast of Major-Major minus Major-Minor and the contrast of Minor-Minor minus Minor-Major during the early and late stages of outcome processing. All activation was thresholded at false discovery rate–corrected $p < 0.05$, with a minimum cluster size of 17 voxels.

**3. RESULTS**

The data from the emotional valence rating for each stimulus type were non-normally distributed. Wilcoxon signed-rank tests revealed that the Major-Major scores were significantly higher than the Major-Minor scores ($p < .001$), and the Minor-Major scores were significantly higher than the Minor-Minor scores ($p < .001$) (Figure 3). These results indicated that the final chords in the stimuli of Major-Major induced more positive emotion in the participants than those of Major-Minor, and those of Minor-Major induced more positive emotion than those of Minor-Minor. In addition, all participants had a 100% correct rate in the post-scan test, which confirmed that the participants were able to discriminate between the four stimulus types.

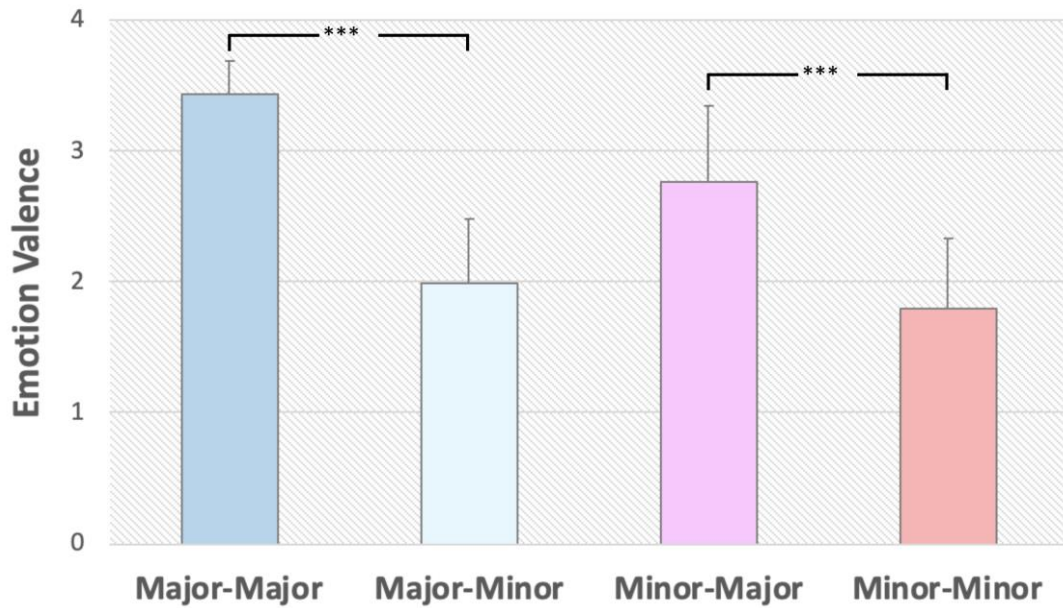

**Figure 3.** Behavioral results of the subjective emotion valence. Error bars indicate standard deviation.
Note: ***$p < 0.001$

The results of the contrasts between Major-Minor and Minor-Minor and between Minor-Major and Minor-Minor during the early stage of outcome processing are summarized in Table 1. Major-Major was associated with significantly greater activity in the subgenual anterior cingulate cortex (sgACC) extending into the vmPFC, posterior middle temporal gyrus, and parahippocampal gyrus than that associated with Major-Minor. No significant difference was found for the contrast of Major-Minor minus Major-Major. Minor-Major was associated with significantly greater activity in the posterior superior temporal gyrus and ventral middle insula than Minor-Minor. No significant difference was found for the contrast of Minor-Minor minus Minor-Major.

**Table 1**. Contrasts between Major-Minor and Minor-Minor and between Minor-Major and Minor-Minor during the early stage of outcome processing ($p < 0.05$, FDR corrected, cluster size > 17).

| Early Stage of Outcome Processing | | | | | |
|---|---|---|---|---|---|
| **Volume information** | **X** | **Y** | **Z** | **t-value** | **Cluster size (voxel)** |
| **Major-Major > Major-Minor** | | | | | |
| Medial orbitofrontal gyrus (ventromedial prefrontal cortex), subgenual anterior cingulate cortex | -12 | 36 | -12 | 6.69 | 61 |
| Middle temporal gyrus | 36 | -60 | 10 | 5.36 | 27 |
| Parahippocampal gyrus | -26 | -44 | -8 | 4.94 | 17 |
| **Major-Minor >Major-Major** | | | | | |
| No significance | | | | | |
| **Minor-Major > Minor-Minor** | | | | | |
| Superior temporal gyrus | 68 | -36 | -2 | 5.19 | 26 |
| Ventral middle insula | 40 | 2 | -16 | 5.10 | 18 |
| **Minor-Minor > Minor-Major** | | | | | |
| No significance | | | | | |

The results of the contrasts between Major-Minor and Minor-Minor and between Minor-Major and Minor-Minor during the late stage of outcome processing are summarized in Table 2. Major-Major was associated with significantly greater activity in the caudate nucleus and superior temporal gyrus than Major-Minor. Conversely, Major-Minor was associated with significantly greater activity in the inferior parietal lobule (IPL), supramarginal gyrus, premotor cortex, dorsolateral prefrontal cortex (dlPFC), inferior frontal gyrus (IFG), pre-supplementary motor area (preSMA) extending into the dorsal anterior cingulate cortex (dACC), caudate nucleus, and posterior superior temporal gyrus, relative to Major-Major. Moreover, Minor-Major was associated with significantly greater activity in the middle occipital gyrus than was Minor-Minor. Conversely, Minor-Minor was associated with significantly greater activity in the ventral anterior insula and supramarginal gyrus (secondary somatosensory cortex; SII) than was Minor-Major.

**Table 2**. Contrasts between Major-Minor and Minor-Minor and between Minor-Major and Minor-Minor during the late stage of outcome processing ($p < 0.05$, FDR corrected, cluster size > 17).

| Late Stage of Outcome Processing | | | | | |
|---|---|---|---|---|---|
| **Volume information** | X | Y | Z | t-value | Cluster size (voxel) |
| **Major-Major > Major-Minor** | | | | | |
| Caudate nucleus | -2 | 20 | 6 | 5.30 | 28 |
| Superior temporal gyrus | 52 | -40 | 12 | 5.22 | 17 |
| **Major-Minor >Major-Major** | | | | | |
| Supramarginal gyrus, inferior parietal lobule | -48 | -52 | 36 | 6.04 | 130 |
| Precentral gyrus, premotor cortex | -38 | 6 | 40 | 5.46 | 68 |
| Inferior frontal gyrus | -38 | 42 | -2 | 5.46 | 17 |
| | -48 | 16 | 2 | 4.84 | 20 |
| Middle frontal gyrus (dorsolateral prefrontal cortex) | -42 | 22 | 46 | 5.04 | 87 |
| Supplementary motor area | -2 | 20 | 56 | 4.65 | 32 |
| Dorsal anterior cingulate cortex | -2 | 30 | 40 | 4.56 | 46 |
| Inferior parietal lobule | -40 | -60 | 52 | 4.5 | 21 |
| **Minor-Major > Minor-Minor** | | | | | |
| Middle occipital gyrus | -22 | -96 | 0 | 5.04 | 31 |
| **Minor-Minor > Minor-Major** | | | | | |
| Ventral anterior insula | 44 | 6 | 6 | 5.64 | 17 |
| Supramarginal gyrus | -62 | -28 | 24 | 5.33 | 23 |
| Ventral anterior insula | -42 | 12 | 0 | 4.76 | 21 |

Contrasts between Major-Minor and Minor-Minor and between Minor-Major and Minor-Minor during the early and late stages of outcome processing are shown in Figure 4. The time courses of focal brain activity in a subset of activated locations, including the anterior insula, vmPFC/sgACC, SII, IPL, and dACC are presented in Figure 5.

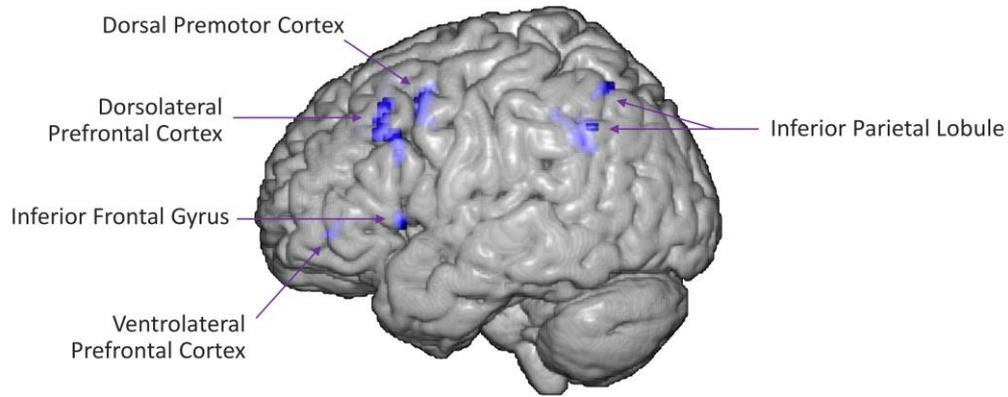
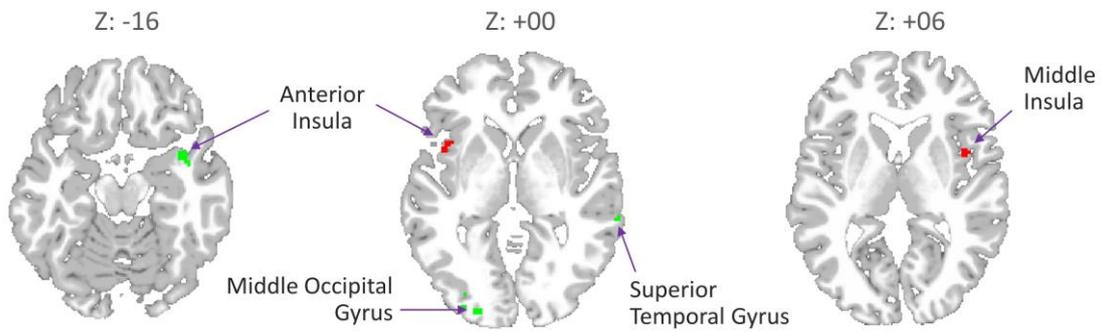

**Figure 4.** Activation map for the contrasts between Major-Minor and Minor-Minor and between Minor-Major and Minor-Minor ($p < 0.05$, FDR corrected, cluster size > 17).

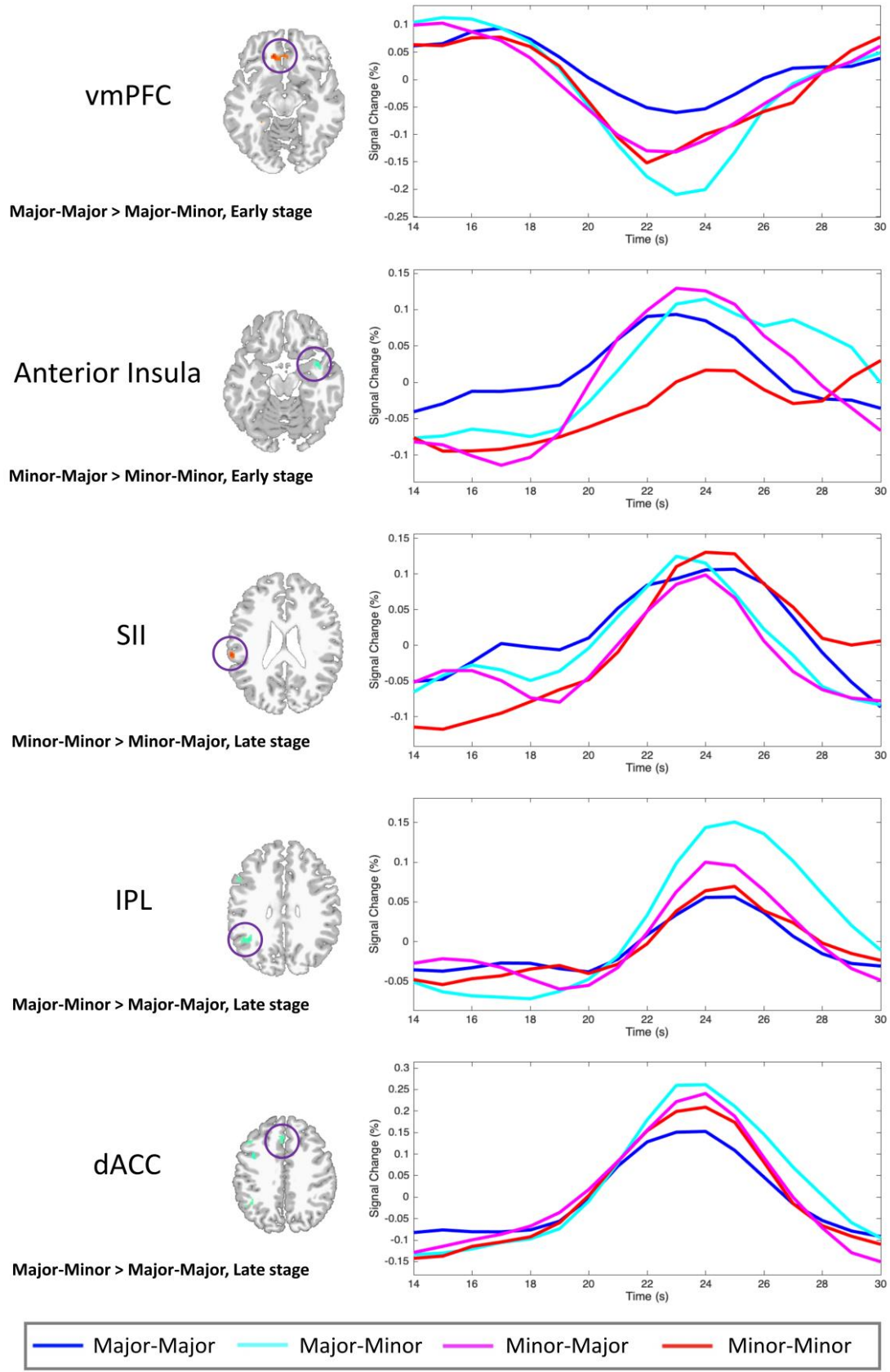

**Figure 5.** Time courses of focal brain activity in a subset of activated locations. Left panels: activation clusters. Right panels: averaged time courses of blood oxygen level–

dependent (BOLD) signal change.
Abbreviations: vmPFC, ventromedial prefrontal cortex; SII, secondary somatosensory cortex; IPL, inferior parietal lobule; dACC, dorsal anterior cingulate cortex.

## 4. DISCUSSION

When listening to music, we tend to predict the reward value of a forthcoming musical event and then experience the feeling arising from the difference between the expected and actual outcome value, namely RPE. The aim of the present fMRI study was the temporal characterization of unsigned and signed RPE components using musical stimuli with mode switch. By manipulating the final chord of J. S. Bach's keyboard pieces, we obtained four types of musical stimuli: Major-Major, Major-Minor, Minor-Minor, and Minor-Major. As expected, the outcome of the final chord in all four stimulus types commonly activated the striatum and ventral anterior insula, which have been found to consistently respond to unsigned RPE (Fouragnan *et al.*, 2018). Regarding signed RPE, the results of the contrasts between Major-Major and Major-Minor replicated and expanded previous findings of the neural correlates of signed RPE, whereas those of the contrasts between Minor-Minor and Minor-Major were unexpected.

During the early stage of outcome processing, the sgACC/vmPFC exhibited significantly greater activity for Major-Major than for Major-Minor. This finding accords with that from the meta-analysis by Fouragnan *et al.* (2018) demonstrating increased activation in the vmPFC for positive RPE relative to negative RPE. In addition, sgACC/vmPFC activity was previously found to correlate with positive musical emotions in the absence of RPE (Trost *et al.*, 2012; Lepping *et al.*, 2016). Therefore, our finding of increased sgACC/vmPFC activity for Major-Major may be linked to both the positive RPE and the positive value of the final tonic chord.

The contrast of Major-Minor minus Major-Major during the late stage of outcome processing yielded activation in a wide network of frontoparietal regions, including preSMA, dACC, ventrolateral prefrontal cortex (vlPFC), dlPFC, dorsal premotor cortex (dPMC), IFG, and IPL / intraparietal sulcus (IPS). This frontoparietal network has been implicated in executive functions such as error/conflict processing and working memory. Overall, this finding is in line with research showing activity increases in frontoparietal regions for chromatic music with intermediate degrees of tonality stability, compared to the music excerpts with high or low degrees of tonality stability (Li *et al.*, 2021). A switch from major to minor mode tends to be associated with a modest decrease of tonality stability. Akin to chromatic music, this switch may trigger a suite of processes related to prediction error. It is well recognized that the dACC and preSMA contribute to monitoring performance in situations involving error or conflict (Mayer *et al.*, 2012; Wessel *et al.*, 2012; Iannaccone *et al.*, 2015). Evidence has highlighted their role in error detection (Fiehler *et al.*, 2004; Orr & Hester, 2012; Gauvin *et al.*, 2016) and action selection (Lau *et al.*, 2006; Akam *et al.*, 2021). In the present study, the participants detected the "error" arising from a minor triad following a major-mode passage and may construct auditory imagery of musical percepts relevant to this minor-triad chord. Moreover, the dACC and preSMA have been implicated in attention control (Peelen *et al.*, 2004; Hilti *et al.*, 2013) and working memory (Manoach *et al.*, 2003; Luo *et al.*, 2014; Walitt *et al.*, 2016). Our finding of greater activity in the dACC/preSMA for Major-Minor than for Major-Major might be related to the attentional selection of the contents of tonal working memory. Musical percepts relevant to this minor triad might be retrieved from long-term memory for the predictive processing of music. This view is compatible with the previous finding that the processing of negative RPE engages arousal-related and motor-preparatory brain structures (Fouragnan *et al.*, 2015; Fouragnan *et al.*, 2018).

Co-activation of the dACC/preSMA and lateral prefrontal cortices is often associated with subsequent response switches (Egner, 2009), but their precise functions are not fully understood. Within the framework of conflict monitoring theory, the dACC signals the need for greater control when conflict arises, and the lateral prefrontal cortices bias attention toward task-relevant processing to ensure the correct response (MacDonald *et al.*, 2000; Botvinick *et al.*, 2001; Weissman *et al.*, 2003). The hierarchical error representation hypothesis states that error signals generated by the medial prefrontal cortices in response to surprising outcomes are used to train representations of expected error in the dlPFC, which are then associated with relevant task stimuli (Alexander & Brown, 2015). Moreover, the left dlPFC and the rostral portion of the dACC were engaged in control processes for response conflict, whereas the left dPMC and caudal portion of dACC were engaged in control processes for perceptual conflict (Kim *et al.*, 2012). In the present study, activation in the dACC/preSMA, left dlPFC, left dPMC, and left IPL for contrast of Major-Minor minus Major-Major suggests that a minor tonic chord following a major-mode passage may lead to auditory conflict, and that the anticipatory auditory imagery of minor-mode music may be associated with response conflict because of the greater uncertainty of minor-mode music than major-mode music (Parncutt, 2014). In addition, our finding of activity in the left IFG and IPL for Major-Minor minus Major-Major agrees with that of Musso *et al.* (2015), who reported increased activity in these regions when a chord violated the rules of harmony or the principles of tonal relatedness. The left IFG found to be activated in the present study was also activated by the negative RPE arising from unexpected noise following a verse-to-chorus transition (anticipatory phase of musical reward) in pop songs (Li *et al.*, 2015). While prior investigations have shown the involvement of the left IFG in conflict processing (Schulte *et al.*, 2009; Chiew & Braver, 2011), additional studies are needed to further elucidate its role in the processing of negative RPE.

We found that activation in the left IPL was more pronounced for Major-Minor versus Major-Major. This result replicated the previous finding that the left IPL preferentially responds to negative RPE (Hauser *et al.*, 2015; Fouragnan *et al.*, 2018) and error/conflict (Orr & Hester, 2012; van de Meerendonk *et al.*, 2013). Kim *et al.* (2010) reported the involvement of the IPL in response and perceptual conflict processing, suggesting that it has a role in orienting attention to the relevant feature of the stimulus. Ciaramelli *et al.* (2010) demonstrated that the left IPL was engaged when participants searched for/anticipated memory targets upon presentation of relevant memory cues, suggesting that it plays a role in allocating top-down attention to memory retrieval. A plethora of prior studies has implicated the IPL/IPS in working memory processes (Foster & Zatorre, 2010; Stevens *et al.*, 2012; Uhlig *et al.*, 2013; Siffredi *et al.*, 2017; Zhu *et al.*, 2020). Following our previous study of chromatic music (Li *et al.*, 2021), we suggest that increased activity in the left IPL/IPS in response to the major-minor switch might reflect an increased working memory load. After exposure to a major-mode passage, listeners' retrieval and maintenance of musical percepts relevant to a final minor tonic chord may place heavier demands on working memory resources than a final major tonic chord. This view is consistent with evidence showing that the left IPL/IPS was activated by phonological working memory tasks (Vigneau *et al.*, 2006), increasing auditory-verbal working memory load (Huang *et al.*, 2013), maintenance of pitch intervals (Tsai *et al.*, 2018), and distantly associated concepts in semantic memory (Ilg *et al.*, 2007). Interestingly, a study of deception found increased activity in the preSMA, left dlPFC, and left IPL during the execution of deception, implicating these areas in the maintenance of the necessary information for deception (Ito *et al.*, 2012). This finding is reminiscent of the *deceptive cadence* in music, which refers to a chord progression where the dominant chord is followed by a chord other than the tonic chord. Although the final chord in Major-Minor was the tonic chord in the parallel minor key and was not the most common final chord of the deceptive cadence (i.e., the minor triad built on the sixth scale degree, or the tonic chord of the relative minor key), the late stage of outcome processing for Major-Minor may also involve executive functions that were needed in the processing of deception (Christ *et al.*, 2009; Ito *et al.*, 2012).

In the present study, the left vlPFC is the most rostral area that showed greater activity for Major-Minor than for Major-Major. This area was not included in the list of neural substrates of the processing of negative RPE in the meta-analysis by Fouragnan *et al.* (2018). Prior research indicated that the left vlPFC displayed significantly increased activation for ambiguous versus unambiguous musical stimuli (Vuust *et al.*, 2011; Li *et al.*, 2021) and linguistic stimuli (Grindrod *et al.*, 2014; Vitello *et al.*, 2014). Moreover, Ruz and Tudela (2011) found that untrustworthy angry partners, compared to trustworthy happy partners, elicited greater activation in the left vlPFC. Based on these studies and our own results, we posit that the increased activation in the left vlPFC for Major-Minor may reflect its role in cognitive control for conflict resolution of remotely related musical percepts in a broad superordinate context of the tonic key, and/or for downregulation of negative emotion (Kanske *et al.*, 2011; Kunz *et al.*, 2011).

Compared to Minor-Minor stimuli, Minor-Major stimuli were associated with significantly increased activity in the right ventral middle insula during the early stage of outcome processing. This region has been found to exhibit greater activity in response to intact music than scrambled music (Fedorenko *et al.*, 2012) or white noise (Fujisawa & Cook, 2011). Moreover, this region preferentially responded to empathy for pleasant affective touch (Lamm *et al.*, 2015) and for music-induced highly pleasant and highly arousing emotions (Trost *et al.*, 2012). In light of these studies, our finding of increased activity in the right ventral middle insula for Minor-Major may reflect the questionnaire result that the participants felt positive emotions at the end of the Minor-Major stimuli.

The contrast of Minor-Minor minus Minor-Major showed activity in the bilateral ventral anterior insula during the late stage of outcome processing. These regions have been typically observed during the experience of pain and empathy for pain (Morrison & Downing, 2007; Mazzola *et al.*, 2009; Mobascher *et al.*, 2009; Straube *et al.*, 2009; Lamm *et al.*, 2011). In addition, activity in the left SII was significantly greater for Minor-Minor than for Minor-Major during the late stage of outcome processing. This region has been reported to respond to laser heat stimuli (Bogdanov *et al.*, 2015), painful heat (Quiton *et al.*, 2014), painful heat with spontaneous facial expressions of pain (Kunz *et al.*, 2011), and own child's sad faces (Kluczniok *et al.*, 2017). We suggest that increased activity in the bilateral ventral anterior insula and the left SII for Minor-Minor may be linked to the participants' negative emotions and feelings.

The activation profiles associated with Minor-Minor and Minor-Major speak against our hypothesis that the contrasts between these two conditions can be explained in terms of the processing of RPE. The activation in the ventral insula and SII that we observed in these contrasts instead lent support to the idea that music plays a role in interoception (Habibi & Damasio, 2014; Koelsch *et al.*, 2021). Furthermore, prior research indicated that, due to its unstable nature and infrequent usage, the minor mode can be regarded as deviated from the normal major mode (Parncutt, 2014). In light of the notion of homeostasis, a reasonable conjecture is that the negative emotional states associated with a minor-mode passage are expected to return to the normal, positive states associated with major-mode music. The unstable, uncertain nature of the minor mode might also relate to the finding that the negative RPE and negative value of the final minor chords were associated with late activation but not early activation. These ideas remain to be tested in future studies.

A major limitation of the present study is that we did not evaluate how the participants expected the final chord in each trial. Even though they reported a low familiarity with the musical stimuli, a knowledge of Baroque music may have allowed them to strongly expect a Picardy cadence in the concluding bars of a minor-mode piece composed by J. S. Bach. Admittedly, we cannot ascertain whether and how the participants utilized that knowledge in the present experiment. Moreover, the RPEs investigated in this study occurred at the final chords of the music pieces. Caution should be exercised in generalizing our results to other music-induced RPEs.

To sum up, the present study was designed to shed light on the neural underpinnings of the cognitive and emotional processing of music-related RPE. Consistent with previous findings, the results of the contrasts between Major-Major and Major-Minor showed that positive RPE was associated with increased activity in the sgACC/vmPFC, whereas negative RPE was associated with increased activity in frontoparietal regions implicated in executive functions. Notably, the contrasts between Minor-Major and Minor-Minor reveal that brain regions implicated in interoception were in charge of the processing of RPE in minor-mode music. This finding may be attributed to the fact that minor-mode music tends to be perceived as more negative and unstable than major-mode music, shedding new light on the emotional connotations of minor mode.